\documentstyle[floats,aps]{revtex}
\begin{document}
\renewcommand{\thefootnote}{\fnsymbol{footnote}}
\draft
\title{\large\bf 
  Graded reflection equation algebras and integrable Kondo impurities 
  in the one-dimensional t-J model}

\author {   Huan-Qiang Zhou \footnote {E-mail: hqz@maths.uq.edu.au}
,  Xiang-Yu Ge \footnote
{E-mail: xg@maths.uq.edu.au},  Jon Links and Mark D. Gould}

\address{Department of Mathematics,University of Queensland,
		     Brisbane, Qld 4072, Australia}

\maketitle

\vspace{10pt}

\begin{abstract}
Integrable Kondo impurities in two cases of the one-dimensional $t-J$ 
model  are studied by means of the boundary ${\bf Z}_2$-graded quantum inverse
scattering method.
The boundary $K$ matrices depending on the local magnetic moments of the
impurities 
are presented  as  nontrivial realizations of the reflection equation
algebras in an impurity Hilbert space.
Furthermore, these  models are  solved by using the algebraic Bethe ansatz 
method
and the Bethe ansatz equations are obtained. 
\end{abstract}




\def\a{\alpha}
\def\b{\beta}
\def\d{\delta}
\def\e{\epsilon}
\def\g{\gamma}
\def\k{\kappa}
\def\l{\lambda}
\def\o{\omega}
\def\t{\theta}
\def\s{\sigma}
\def\D{\Delta}
\def\L{\Lambda}


\def\beq{\begin{equation}}
\def\eeq{\end{equation}}
\def\bea{\begin{eqnarray}}
\def\eea{\end{eqnarray}}
\def\ba{\begin{array}}
\def\ea{\end{array}}
\def\no{\nonumber}
\def\le{\langle}
\def\re{\rangle}
\def\lt{\left}
\def\rt{\right}

\newcommand{\sect}[1]{\setcounter{equation}{0}\section{#1}}
\renewcommand{\theequation}{\thesection.\arabic{equation}}
\newcommand{\reff}[1]{eq.~(\ref{#1})}

\vskip.3in

\sect{Introduction\label{int}}

The Kondo problem describing the effect due to the exchange interaction
between the magnetic impurity and the conduction electrons plays a
very important role in condensed matter physics \cite {Kon64}. Wilson
\cite {Wil75} developed
a very powerful numerical renormalization group approach, and the
model was also solved by the coordinate Bethe ansatz method \cite 
{Wie83,And83} which
gives the specific heat and magnetization. More recently, a conformal
field theory approach was developed by Affleck and Ludwig \cite {AL91} based on 
a work by Nozi{\`e}res \cite {Noz74}. In the conventional Kondo problem,
the
interaction between conduction electrons is discarded, due to the
fact that the interacting electron system can be described as a Fermi
liquid. Recently there has been substantial research  devoted to the
investigation
of the theory of impurities coupled to  Luttinger liquids. 
Such a problem was first considered by Lee and Toner \cite {LT92}. By using the
perturbative renormalization group theory they found that the 
Kondo temperature crosses from a generic power law dependence on 
the Kondo coupling
constant to an exponential one in the infinite limit.  Afterwards, a  ``poor 
man's'' scaling  procedure was
carried out by Furusaki and Nagaosa \cite {FN94}, who found
a stable strong coupling fixed point  for both antiferromagnetic
and ferromagnetic cases. On the other hand, boundary conformal field
theory predicts two types of critical behaviours, i.e., either a local 
Fermi liquid with standard low-temperature thermodynamics or the non-Fermi
liquid observed by Furusaki and Nagaosa \cite {FN94}. However, in order 
to get a full
picture about the critical behaviour
of Kondo impurities coupled to Luttinger liquids, some simple
integrable models
which allow exact solutions are desirable.

Several integrable magnetic or nonmagnetic impurity problems
describing impurities embedded
in systems of correlated electrons have so far appeared in the
literature. Among them are  versions of the supersymmetric $t-J$
model with impurities \cite {BAR94,BEF97,SZ,LF98,AR}. 
Such an idea to incorporate an impurity into a
closed chain  dates back to Andrei and Johanesson \cite {AJ84} (see
also \cite {LS,ZJ89}). However, the model
thus constructed suffers from the lack of backward scattering and results in
a very complicated Hamiltonian which is difficult to be justified
on physical grounds. Therefore, as observed by Kane and
Fisher \cite {KF92}, it is advantageous to adopt open boundary
conditions with
the impurities situated at the ends of the chain 
when studying  Kondo impurities coupled to
integrable strongly correlated electron systems \cite {PW97,WV,ZG}. 

In this paper, integrable Kondo impurities with arbitrary spin coupled
to the one-dimensional $t-J$ open chain are constructed following a formalism
recently advocated in \cite {ZG}. Our 
new input is to search for integrable boundary $K$ matrices
depending on the local magnetic moments of impurities, 
which arise as a nontrivial realization of the ${\bf Z}_2$-graded 
reflection equation (RE) algebras in a finite dimensional quantum
space,which may be interpretated as an impurity Hilbert space.
It should be emphasized that our new non-c-number boundary 
$K$ matrices are highly nontrivial, 
in the sense that they can not be factorized into the product of a
c-number boundary $K$ matrix and the corresponding local monodromy
matrices. The models we present are  solved by means of the 
algebraic Bethe ansatz method and
the Bethe ansatz equations are derived.

The layout of this paper is the following.  We begin by reviewing the
${\bf Z}_2$-graded boundary Quantum Inverse Scattering Method (QISM)  as
formulated in \cite{Zhou97}. We then introduce two integrable
cases of the one-dimensional $t-J$ model with Kondo impurities on the
boundaries. Integrability of the models is established by relating the
Hamiltonians to one parameter families of commuting transfer matrices.
This is achieved through solving the reflection equations for 
non-c-number solutions. 
Finally we solve the model by means of the algebraic 
Bethe ansatz method and derive the Bethe ansatz equations.

\sect{Graded Reflection Equation Algebra and Transfer Matrix 
      \label{for}}

In this section, we give a brief review about
the ${\bf Z}_2$-graded boundary quantum inverse scattering method.
To begin, let
$V$ be a finite-dimensional linear superspace. Let
$R\in End(V\otimes V)$ be a solution to the ${\bf Z}_2$-graded quantum
Yang-Baxter equation (QYBE)   
\beq
R_{12}(u_1-u_2)R_{13}(u_1-u_3)R_{23}(u_2-u_3)=R_{23}(u_2-u_3)
R_{13}(u_1-u_3)R_{12}(u_1-u_2).
\eeq
Here $R_{jk}(u)$ denotes the matrix on $V\otimes V\otimes V$ acting on the
$j$-th and $k$-th superspaces and as an identity on the remaining 
superspace. The
variables $u_1,~u_2$ and $u_3$ are  spectral parameters. The 
tensor product
should be understood in the graded sense, that is the
multiplication rule for any homogeneous elements $x,\,y,\,x',\,y'\in End
V$ is given by
\beq
(x\otimes y) (x'\otimes y')=(-1)^{[y][x']}\;(xx'\otimes yy'),
\label{rule} \eeq
where $[x]$ stands for the ${\bf Z}_2$-grading of the element $x$. Let $P$
be the ${\bf Z}_2$-graded permutation operator in $V\otimes V$. Then
$P(x\otimes y)=(-1)^{[x][y]}y\otimes x,~\forall x,y\in V$ and
$R_{21}(u)=P_{12}R_{12}(u)P_{12}$. 

We form the monodromy matrix
$T(u)$ for a $L$-site lattice  chain by  
\beq
T(u)=R_{0L}(u)\cdots R_{01}(u).   
\eeq
Indeed, one may
show that $T(u)$ generates  a representation of the ${\bf Z}_2$-graded 
quantum Yang-Baxter algebra,
\beq
R_{12}(u_1-u_2)\stackrel {1} {T}(u_1) \stackrel{2}{T}(u_2)
  =\stackrel{2}{T}(u_2)\stackrel{1}{T}(u_1) R_{12}(u_1-u_2),\label{rtt-ttr}
\eeq
where for notational convenience we have 
$$\stackrel {1} {T}(u)=T_{13}(u),~~~\stackrel{2}{T}(u)=T_{23}(u), $$ 
and the subscript 3 now labels the quantum superspace $W=V^{\otimes
L}$. 

In order to describe integrable Kondo impurities in strongly correlated
electronic models with open boundary
conditions, we need to introduce an appropriate ${\bf Z}_2$-graded 
reflection equation (RE) algebra. 
We introduce the associative superalgebras ${\cal T}_-$ and ${\cal T}_+$
defined by the $R$-matrix and the relations
\bea
&&R_{12}(u_1-u_2)\stackrel {1} {\cal T}_-(u_1) R_{21}(u_1+u_2)
\stackrel {2}{\cal T}_-(u_2)= 
\stackrel {2}{\cal T}_-(u_2) R_{12}(u_1+u_2)
\stackrel {1}{\cal T}_-(u_1) R_{21}(u_1-u_2),\no\\
&&R_{21}^{st_1 \; ist_2}(-u_1+u_2)\stackrel {1}{{\cal T}^{st_1}_+}(u_1) 
{ R}_{12}(-u_1-u_2+\eta)
\stackrel {2}{{\cal T}^{ist_2}_+}(u_2)\no\\
&&~~~~~~~~~~~~~~~~~~~~~~=\stackrel {2}{{\cal T}^{ist_2}_+}(u_2)  R_{21}(-u_1-u_2+\eta)
\stackrel {1}{{\cal T}^{st_1}_+}(u_1) R_{12}^{st_1\; ist_2}(u_1-u_2). 
\label{re-alg}    
\eea
Here $\eta$ is the so-called crossing parameter and
$st_i$ stands for the supertransposition taken in the $i$-th space,
whereas $ist_i$  is the inverse operation of $st_i$. 
One of the important steps towards correctly formulating 
the ${\bf Z}_2$-graded case is to introduce
in the second RE in (\ref{REs})
the inverse operation of the supertransposition.
In our  cases, the $R$-matrices enjoy the unitarity property,
\beq
 R_{12}(u_1-u_2)R_{21}(-u_1+u_2) = 1,\label{II6}
\eeq
and the crossing unitarity
\beq
 R^{st_1}_{12}(u_1-u_2)R^{st_1}_{21}(-u_1+u_2+\eta) = 1.\label{II7}
\eeq

One can obtain a class of realizations of the superalgebras ${\cal T}_+$  and
${\cal T}_-$  by choosing  ${\cal T}_{\pm}(u)$ to be the form
\beq
{\cal T}_-(u) = T_-(u) \tilde {\cal T}_-(u) T^{-1}_-(-u),~~~~~~ 
{\cal T}^{st}_+(u) = T^{st}_+(u) \tilde {\cal T}^{st}_+(u) 
  \lt(T^{-1}_+(-u)\rt)^{st},\label{t-,t+} 
\eeq
with
\beq
T_-(u) = R_{0M}(u) \cdots R_{01}(u),~~~~
T_+(u) = R_{0L}(u) \cdots R_{0,M+1}(u),~~~~ 
\tilde {\cal T}_{\pm}(u) = K_{\pm}(u),
\eeq
where $M$ is any index between 1 and $L$, and 
$K_{\pm}(u)$, called boundary $K$-matrices, 
are representations of  ${\cal T}_{\pm}  $. 
 In the following, without loss of generality, we shall choose
$M=L$ so that ${\cal T}_+(u)\equiv K_+(u)$.

The $K$-matrices $K_\pm(u)$ satisfy the same
relations as ${\cal T}_\pm(u)$, respectively. That is the $K$-matrices
obey the following REs
\bea
&&R_{12}(u_1-u_2)\stackrel {1} {K}_-(u_1) R_{21}(u_1+u_2)
\stackrel {2}{K}_-(u_2)= 
\stackrel {2}{K}_-(u_2) R_{12}(u_1+u_2)
\stackrel {1}{K}_-(u_1) R_{21}(u_1-u_2),\no\\ 
&&R_{21}^{st_1 \; ist_2}(-u_1+u_2)\stackrel {1}{K^{st_1}_+}(u_1) 
R_{12}(-u_1-u_2+\eta)
\stackrel {2}{K^{ist_2}_+}(u_2)\no\\ 
&&~~~~~~~~~~~~~~~~~~~=\stackrel {2}{K^{ist_2}_+}(u_2) R_{21}(-u_1-u_2+\eta)
\stackrel {1}{K^{st_1}_+}(u_1) R_{12}^{st_1\; ist_2}(-u_1+u_2),\label{REs}    
\eea

Following  Sklyanin's approach \cite{Skl88},
one defines the boundary transfer matrix $\tau(u)$ as
\beq
\tau(u) = str (K_+(u){\cal T}_-(u))=str\lt(K_+(u)T(u)K_-(u)T^{-1}(-u)\rt).
\eeq
Then it can be shown that \cite{Zhou97}  
\beq
[\tau(u_1),\tau(u_2)] = 0.
\eeq
Although many attempts have been made to find c-number
boundary $K$ matrices, which may be referred to as the fundamental
representation, it is no doubt very interesting to search for
non-c-number $K$ matrices arising as representations in some Hilbert spaces,
which may be interpreted as impurity  spaces.

\sect{Integrable non-c-number boundary $K$-matrices and 
      Kondo impurities in the one-dimensional $t-J$ model \label{Boun}}

Let $c_{j,\s}^\dagger$ and $c_{j,\s}$ denote  creation and
annihilation operators for  conduction electrons with spin $\s$ at
site $j$, satisfying the anti-commutation relations given by
$\{c_{i,\s}^\dagger, c_{j,\s'}\}=\d_{ij}\d_{\s\s'}$, where 
$i,j=1,2,\cdots,L$ and $\s,\s'=\uparrow,\;\downarrow$. We consider the
following type of Hamiltonians describing two magnetic impurities coupled to
 open t-J chains
\bea
H&=&-t\sum _{j=1, \s}^{L-1}{\cal P} (c^\dagger_{j\s}c_{j+1\s}+H. c.){\cal
P}+
   \sum_{j=1}^{L-1}(J{\bf S}_j \cdot {\bf S}_{j+1} + 
  Vn_jn_{j+1})+\no\\
& &J_a{\bf S}_1 \cdot {\bf S}_a+V_an_1+J_b{\bf S}_L \cdot {\bf S}_b
+V_b n_L.\label {ham1}
\eea
Above,  the projector ${\cal P}=\prod ^L_{j=1} (1-n_{j\uparrow}n_{j\downarrow})$
ensures that double electron 
occupancies of sites are forbidden; $J_\a,V_\a (\a=a,b)$ 
are the Kondo coupling constants and the
impurity scalar potentials respectively; ${\bf S_j}$ as usual is the
vector spin operator for   
 the conduction electrons at site $j$; ${\bf S}_{\a} (\a = a,b)$ are the local
moments with spin-$s$ located at the left and right ends of
the system respectively;
 $n_{j\s}$ is the number density operator
$n_{j\s}=c_{j\s}^{\dagger}c_{j\s}$,
$n_j=n_{j\uparrow}+n_{j\downarrow}$.

For the choices 
\beq t=1,~~~\,J=2,~~~\, V=-\frac{1}{2}, \label{con1}   \eeq    
it has been shown in 
refs. \cite {SCH87,SAK91,BBO91} that the bulk Hamiltonian acquires an
underlying supersymmetry algebra given by $gl(2|1)$ in the minimal
representation. Throughout we will refer to this case as the
supersymmetric t-J model. Integrability of this model 
on a closed chain
with periodic boundary conditions was established independently in works by 
by Essler and Korepin \cite {EK92} and Foerster and Karowski
\cite {FK92} by showing that the model can be constructed using the
QISM. Furthermore, open chain integrability with appropriate
boundary conditions was shown in refs. \cite{FK93a,Gon94a,E}.  
 It is quite interesting to note that although
the introduction  of integrable  impurities we propose below 
spoils the supersymmetry, there 
still remains  $su (2)$ symmetry in the Hamiltonian (\ref {ham1}) which
maintains conservation of total spin and electron number.
We will establish the quantum integrability of the Hamiltonian
(\ref{ham1}) for the special choice of the model parameters (\ref{con1})
and  
\beq
J_\a = -\frac {2}{(c_\a -s)(c_\a+s+1)},
V_\a = -\frac {c_\a^2-s(s+1)}{(c_\a -s)(c_\a+s+1)}.
\label{con3}   \eeq
This is achieved by showing that this model can be derived from
the QISM.  
 Our result is consistent with the applicability of the
coordinate Bethe ansatz method discussed in \cite {PW97}.

Another choice of couplings which leads to an integrable model on the
closed periodic chain is given 
by 
\beq t=1,~~~\,J=-2,~~~\, V=-\frac{3}{2}, \label{con2} \eeq     
 as shown by Schlottmann \cite{SCH87} corresponding to an $su(3)$
 invariant solution of the Yang-Baxter equation.     
In this case we can introduce integrable Kondo impurities on the
boundary by choosing  
\beq
J_\a = \frac {2}{(c_\a +s)(c_\a-s-1)},
V_\a = \frac {c_\a^2-1-s(s+1)}{(c_\a +s)(c_\a-s-1)}.
\label{con4}  \eeq
Below, we describe how the aforementioned integrable cases are obtained.

Let us recall that the local Hamiltonian of
the  supersymmetric $t-J$  model is derived from an $R$-matrix
satisfying the Yang-Baxter equation which has the form \cite{EK92,FK92} 
\beq
R(u)= uI+ P \equiv 
\left ( \begin {array} {ccccccccc}
u-1&0&0&0&0&0&0&0&0\\
0&u&0&-1&0&0&0&0&0 \\
0&0&u&0&0&0&1&0&0\\
0&-1&0&u&0&0&0&0&0\\
0&0&0&0&u-1&0&0&0&0\\
0&0&0&0&0&u&0&1&0\\
0&0&-1&0&0&0&u&0&0\\
0&0&0&0&0&1&0&u&0\\
0&0&0&0&0&0&0&0&u+1
\end {array}  \right ),\label{r1}
\eeq
where $u$ is the spectral parameter and 
$$P=\sum_{ij}(-1)^{[j]} e^i_j\otimes e^j_i$$ 
is the form of  the ${\bf  Z}_2$-graded permutation
operator in accordance with the rule (\ref{rule}). 
We chose to adopt the ${\bf Z}_2$-grading 
$[1]=[2]=1,\,[3]=0$ on the indices labelling the basis vectors. 

We now solve  (\ref{REs}) 
for $K_+(u)$ and $K_-(u)$. For the quantum $R$-matrix (\ref {r1}),
One may 
check that the matrix $K_-(u)$ given by
\beq
K_-(u)=   \left ( \begin {array}
{ccc}
A_-(u)&B_-(u)&0\\
C_-(u)&D_-(u)&0\\
0&0&1
\end {array} \right ),\label{K}
\eeq
where
\bea
A_-(u)&=&\frac {c_a(c_a+1)-u^2+u-s_a(s_a+1)+2u {\bf S}^z_a}
{(c_a+u-s_a)(c_a+u+s_a+1)},\no\\
B_-(u)&=&\frac {2u {\bf S}^-_a}
{(c_a+u-s_a)(c_a+u+s_a+1)},\no\\
C_-(u)&=&\frac {2u {\bf S}^+_a}
{(c_a+u-s_a)(c_a+u+s_a+1)},\no\\
D_-(u)&=&\frac {c_a(c_a+1)-u^2+u-s_a(s_a+1)-2u {\bf S}^z_a}
{(c_a+u-s_a)(c_a+u+s_a+1)},
\eea
satisfies the first equation of (\ref{REs})(for the details, see
the Appendix). Here ${\bf S}^{\pm}={\bf S}^x \pm
i{\bf S}^y$.
The matrix $K_+(u)$ can be obtained from the isomorphism of the
superalgebras  ${\cal T}_-  $ and ${\cal T}_+  $. Indeed, given a solution
$K_- $ of the first equation of (\ref{REs}), then $K_+(u)$ defined by
\beq
 K_+^{st}(u) =   K_-(-u+\frac {1}{2})\label{t+t-1}
\eeq
is a solution of the second equation of (\ref{REs}). 
The proof follows from some algebraic computations upon
substituting (\ref{t+t-1}) into  
the second equation of (\ref{REs}) and making use
of the properties (\ref {II6}) and (\ref {II7}) of the $R$-matrix with
$\eta =1$.
Therefore, one may choose the boundary matrix $K_+(u)$ as 
\beq
K_+(u)=   \left ( \begin {array}
{ccc}
A_+(u)&B_+(u)&0\\
C_+(u)&D_+(u)&0\\
0&0&1 
\end {array} \right ),\label{k+}
\eeq
where
\bea
A_+(u)&=&\frac {c_b^2-u^2-s_b(s_b+1) +(2u-1) {\bf S}^z_b}
{(c_b+u+s_b)(c_b+u-s_b-1)},\no\\
B_+(u)&=&\frac {(2u-1){\bf S}^-_b}
{(c_b+u+s_b)(c_b+u-s_b-1)},\no\\
C_+(u)&=&\frac {(2u-1){\bf S}^+_b}
{(c_b+u+s_b)(c_b+u-s_b-1)},\no\\
D_+(u)&=&\frac {c_b^2-u^2-s_b(s_b+1) -(2u-1) {\bf S}^z_b}
{(c_b+u+s_b)(c_b+u-s_b-1)}.
\eea

It can be shown  that the 
Hamiltonian (\ref{ham1}) is related to the logarithmic derivative of the
transfer matrix
$\tau (u)$ with respect to the spectral parameter $u$ at $u=0$ (up to 
an additive chemical potential term)
\beq
 -H= \sum _{j=1}^{L-1} h_{j,j+1} + \frac {1}{2} \stackrel {1}{K'}_-(0)
+\frac {str_0 K_+(0)H_{L0}}{str_0 K_+(0)},
\label{HAM} \eeq
with 
$$h=\frac{d}{du}PR(u). $$ 
For this case we obtain (\ref{ham1}) subject to the constraints
(\ref{con1}, \ref{con3}). This implies that this model admits
an infinite number
of conserved currents  thus assuring integrability.

The second choice of integrable couplings results from use of 
an $R$-matrix  obtained by imposing ${\bf Z}_2$-grading to the
fundamental $su(3)$ $R$-matrix and which reads 
\beq
R(u)=\left ( \begin {array} {ccccccccc}
-u-1&0&0&0&0&0&0&0&0\\
0&-u&0&-1&0&0&0&0&0 \\
0&0&u&0&0&0&1&0&0\\
0&-1&0&-u&0&0&0&0&0\\
0&0&0&0&-u-1&0&0&0&0\\
0&0&0&0&0&u&0&1&0\\
0&0&-1&0&0&0&u&0&0\\
0&0&0&0&0&1&0&u&0\\
0&0&0&0&0&0&0&0&u+1
\end {array}  \right ),\label{r2}
\eeq
where again $u$ is the spectral parameter and we adopt the same choice
for the ${\bf Z}_2$-grading of the basis states as before.   
 
 We now solve (\ref{REs})
 for $K_+(u)$ and $K_-(u)$ for this $R$-matrix (\ref {r2}).  
 One may check that the matrix $K_-(u)$ given by
 \beq
 K_-(u)=   \left ( \begin {array}
 {ccc}
 A_-(u)&B_-(u)&0\\
 C_-(u)&D_-(u)&0\\
 0&0&1
 \end {array} \right ),\label{k1-}
 \eeq
 where
 \bea
 A_-(u)&=&\frac {c_a(c_a-1)-u^2-u-s_a(s_a+1)-2u {\bf S}^z_a}
 {(c_a-u+s_a)(c_a-u-s_a-1)},\no\\
 B_-(u)&=&-\frac {2u {\bf S}^-_a}
 {(c_a-u+s_a)(c_a-u-s_a-1)},\no\\
 C_-(u)&=&-\frac {2u {\bf S}^+_a}
 {(c_a-u+s_a)(c_a-u-s_a-1)},\no\\
 D_-(u)&=&\frac {c_a(c_a-1)-u^2-u-s_a(s_a+1)+2u {\bf S}^z_a}
 {(c_a-u+s_a)(c_a-u-s_a-1)},
 \eea
 satisfies the first equation of (\ref{REs}). 
 For this case $K_+(u)$ defined by
  \beq
  K_+^{st}(u) =  K_-(-u-\frac {3}{2})\label{t+t-2}
  \eeq
  is a solution of the second equation of (\ref{REs}), since the
  crossing parameter $\eta =-3$.
  Therefore, we may choose the boundary matrix $K_+(u)$ as
  \beq
  K_+(u)=   \left ( \begin {array}
  {ccc}
  A_+(u)&B_+(u)&0\\
  C_+(u)&D_+(u)&0\\
  0&0&1
  \end {array} \right ),\label{k1+}
  \eeq
  where
  \bea
  A_+(u)&=&-\frac {c_b^2-u^2-2u-s_b(s_b+1)-1 -(2u+3) {\bf S}^z_b}
  {(c_b-u+s_b-1)(c_b-u-s_b-2)},\no\\
  B_+(u)&=&\frac {(2u+3){\bf S}^-_b}
  {(c_b-u+s_b-1)(c_b-u-s_b-2)},\no\\
  C_+(u)&=&\frac {(2u+3){\bf S}^+_b}
  {(c_b-u+s_b-1)(c_b-u-s_b-2)},\no\\
  D_+(u)&=&-\frac {c_b^2-u^2-2u-s_b(s_b+1)-1 +(2u+3) {\bf S}^z_b}
  {(c_b-u+s_b-1)(c_b-u-s_b-2)}.
  \eea

As before, the Hamiltonian in this case is defined as the logarithmic
derivative of the transfer matrix at zero spectral parameter giving the
same general form (\ref{ham1}) subject to the constraints (\ref{con2},
\ref{con4}). 

Before concluding this subsection,we would like to point out that 
the quantum integrability of other two  cases corresponding to the 
same $t$ but with negation of all
$J$'s and $V$'s follows from the transformation
$c^\dagger _j \rightarrow (-1)^j c^\dagger _j$,
$c _j \rightarrow (-1)^j c _j$.

\sect{The Bethe ansatz solutions \label{bethe}}

Having established the quantum integrability of the models, let us
first
diagonalize the Hamiltonian (\ref {ham1})
by means of the algebraic Bethe ansatz method
\cite {Skl88,Gon94} for the choice of couplings (\ref{con1},
\ref{con3}).
We  introduce the `doubled' monodromy matrix $U(u)$,
\beq
U(u)=T(u)K_-(u)\tilde{T}(u) \equiv
 \left ( \begin {array}
{ccc}
{\cal A}_{11}(u)&{\cal A}_{12}(u)&{\cal B}_1(u)\\
{\cal A}_{21}(u)&{\cal A}_{22}(u)&{\cal B}_2(u)\\
{\cal C}_1(u)&{\cal C}_2 (u)& {\cal D}(u)
\end {array} \right ),\label{u1}
\eeq
where $\tilde {T}(u)=T^{-1}(-u)$.
Substituting into the graded reflection equation (\ref {REs}) ,
we may draw the following commutation relations,
\bea
{\check {\cal A}}_{bd}(u_1){\cal C}_c(u_2)&=&\frac {(u_1-u_2-1)(u_1+u_2)}
{(u_1-u_2)(u_1+u_2+1)}r(u_1+u_2+1)^{eb}_{gh}r(u_1-u_2)^{ih}_{cd}
{\cal C}_e(u_2){\check {\cal A}}_{gi}(u_1)-\no\\
& &\frac {4u_1u_2}{(u_1+u_2+1)(2u_1+1)(2u_2+1)}r(2u_1+1)^{gb}_{cd}
{\cal C}_g(u_1) {\cal D}(u_2) + \no\\
& & \frac {2u_1}{(u_1-u_2)(2u_1+1)}
r(2u_1+1)^{gb}_{id} {\cal C}_g (u_1) {\check {\cal A}}_{ic}(u_2),\label
{cr1}\\
{\cal D}(u_1){\cal C}_b(u_2) 
&=&\frac {(u_1-u_2-1)(u_1+u_2)}
{(u_1-u_2)(u_1+u_2+1)}
{\cal C}_b(u_2){\cal D}(u_1)+\frac {2u_2}{(u_1-u_2)(2u_2+1)}
{\cal C}_b(u_1){\cal D}(u_2)\no\\
& &  -\frac {1}{u_1+u_2+1}{\cal C}_d(u_1){\check {\cal A}}_{db}(u_2).
\eea
Here $ {\cal A} _{bd}(u) = {\check {\cal A}}_{bd}(u) + \frac {1}{2u+1}
\delta _{bd}{\cal D}(u)$ and the matrix $r(u)$, 
which in turn satisfies the quantum Yang-Baxter
equation, takes the form,
\beq
r^{bb}_{bb}(u)=1,~~~~~r^{bd}_{bd}=-\frac
{1}{u-1},~~~~~r^{bd}_{db}(u)=\frac {u}{u-1},(b \neq d,b,d = 1,2).
\eeq
Choosing the Bethe state $|\Omega \rangle $ as
\beq
|\Omega \rangle = {\cal C}_{i_1}(u_1) \cdots {\cal
C}_{i_N}(u_N)|\Psi\rangle F^{i_1\cdots i_N},
\eeq
with $|\Psi\rangle $ being the pseudovacuum, and applying the transfer
matrix $\tau (u)$
to the state $|\Omega\rangle$, we have
$\tau (u) |\Omega \rangle =\Lambda(u) |\Omega \rangle$, with the
eigenvalue,
\bea
\Lambda (u)&=& \frac {2u-1}{2u+1}\frac {(c_b+u-s_b)}{(c_b+u+s_b)}
\frac {(c_b+u+s_b+1)}{(c_b+u-s_b-1)}
(-\frac {u+1}{u-1})^L
\prod ^N_{j=1} \frac {(u+u_j)(u-u_j-1)}{(u-u_j)(u+u_j+1)}\no\\
& &-\frac {2u}{2u+1} (-\frac {u^2}{u^2-1})^L 
\prod ^N_{j=1} \frac {(u+u_j)(u-u_j-1)}{(u-u_j)(u+u_j+1)}
\Lambda ^{(1)}(u;\{u_i\}),
\eea
provided the parameters $\{ u_j\}$ satisfy
\beq
 \frac {2u_j-1}{2u_j}\frac {(c_b+u_j-s_b)}{(c_b+u_j+s_b)}
\frac {(c_b+u_j+s_b+1)}{(c_b+u_j-s_b-1)}
(\frac {u_j+1}{u_j})^{2L}=\Lambda
^{(1)}(u_j;\{u_i\}). \label {bethe11}
\eeq
Here $\Lambda ^{(1)}(u;\{u_i\})$ is the eigenvalue of the transfer
matrix $\tau ^{(1)}(u)$ for the reduced problem, which arises out of the
$r(u)$ matrices from the first term in the right hand side of (\ref {cr1}),
with
the reduced boundary K matrices $K_{\pm}^{(1)}(u)$ as,
\beq
K^{(1)}_-(u)=
  \left ( \begin {array}
{cc}
A^{(1)}_-(u)&B^{(1)}_-(u)\\
C^{(1)}_-(u)&D^{(1)}_-(u)
\end {array} \right ),\label{k0-}
\eeq
where
\bea
A^{(1)}_-(u)&=&\frac {c_a^{2}-u^2-s_a(s_a+1)+(2u+1) {\bf S}^z_a}
{(c_a+u-s_a)(c_a+u+s_a+1)},\no\\
B^{(1)}_-(u)&=&\frac {(2u+1) {\bf S}^-_a}
{(c_a+u-s_a)(c_a+u+s_a+1)},\no\\
C^{(1)}_-(u)&=&\frac {(2u+1) {\bf S}^+_a}
{(c_a+u-s_a)(c_a+u+s_a+1)},\no\\
D^{(1)}_-(u)&=&\frac {c_a^{2}-u^2-s_a(s_a+1)-(2u+1) {\bf S}^z_a}
{(c_a+u-s_a)(c_a+u+s_a+1)},
\eea
and
\beq
K^{(1)}_+(u)=
  \left ( \begin {array}
{cc}
A^{(1)}_+(u)&B^{(1)}_+(u)\\
C^{(1)}_+(u)&D^{(1)}_+(u)
\end {array} \right ),\label{k0+}
\eeq
where
\bea
A^{(1)}_+(u)&=&\frac {c_b^2-u^2-s_b(s_b+1) +(2u-1) {\bf S}^z_b}
{(c_b+u+s_b)(c_b+u-s_b-1)},\no\\
B^{(1)}_+(u)&=&\frac {(2u-1){\bf S}^-_b}
{(c_b+u+s_b)(c_b+u-s_b-1)},\no\\
C^{(1)}_+(u)&=&\frac {(2u-1){\bf S}^+_b}
{(c_b+u+s_b)(c_b+u-s_b-1)},\no\\
D^{(1)}_+(u)&=&\frac {c_b^2-u^2-s_b(s_b+1) -(2u-1) {\bf S}^z_b}
{(c_b+u+s_b)(c_b+u-s_b-1)}.
\eea
Here $K^{(1)}_- (u)$, the boundary K matrices after  the first
nesting, follows from the relations,
\bea
{\check {\cal A}}_{dd}(u)|\Psi\rangle&\equiv&
\frac {2u}{2u+1}K^{(1)}_{dd}(u)|\Psi \rangle =
(K_-(u)_{dd}-\frac{1}{2u+1})
(-\frac{u^2}{u^2-1})^L|\Psi\rangle,\no\\
{\check {\cal A}}_{db}(u)|\Psi\rangle&\equiv&
\frac {2u}{2u+1}K^{(1)}_{db}(u)|\Psi \rangle=
K_-(u)_{db} (-\frac{u^2}{u^2-1})^L|\Psi\rangle.
\eea
Indeed, applying the monodromy matrix
$T(u)$ and its ``adjoint'' ${\tilde T}(u)$ to the pseudovacuum, we have
\bea
T_{dd}(u)|\Psi\rangle&=&u^L|\Psi\rangle,~~~
T_{33}(u)|\Psi\rangle=(u+1)^L|\Psi\rangle,\no\\
T_{3d}(u)|\Psi\rangle&\neq& 0,~~~
T_{db}(u)|\Psi\rangle=0,~~~
T_{d3}(u)|\Psi\rangle= 0,\no\\
\tilde{T}_{dd}(u)|\Psi\rangle&=&(-\frac{u}{u^2-1})^L|\Psi\rangle,~~~
\tilde{T}_{33}(u)|\Psi\rangle=(-\frac{u+1}{u^2-1})^L|\Psi\rangle,\no\\
\tilde{T}_{3d}(u)|\Psi\rangle&\neq& 0,~~~
\tilde{T}_{db}(u)|\Psi\rangle=0,~~~
\tilde{T}_{d3}(u)|\Psi\rangle=0.
\eea
Then we have 
\bea
{\cal D}(u)|\Psi\rangle&=&
(-\frac{u+1}{u-1})^L|\Psi\rangle,\no\\
{\cal B}_d(u)|\Psi\rangle&=&0,~~~~~~
{\cal C}_d(u)|\Psi\rangle\neq 0,\no\\
{\cal A}_{db}(u)|\Psi\rangle&=&
(-\frac{u^2}{u^2-1})^L K_-(u)_{db}|\Psi\rangle,\no\\
{\cal A}_{dd}(u)|\Psi\rangle&=&
(-\frac {u^2}{u^2-1})^L (K_-(u)_{dd}-\frac{1}{2u+1})|\Psi\rangle
+\frac{1}{2u+1}(-\frac{u+1}{u-1})^L|\Psi\rangle.
\eea
In our calculation, use of the following 
relations has also been made  
\bea
(2u+1)T_{13}(u)\tilde{T}_{31}(u)+T_{11}(u)\tilde{T}_{11}+
T_{12}(u)\tilde{T}_{21}(u)&=&-(2u-1)\tilde{T}_{31}T_{13}(u)
+\tilde{T}_{32}T_{23}(u)
+\tilde{T}_{33}T_{33}(u),\no\\
(2u+1)T_{13}(u)\tilde{T}_{32}(u)+T_{11}(u)\tilde{T}_{12}
+T_{12}(u)\tilde{T}_{22}(u)&=&-2u\tilde{T}_{32}T_{13}(u),\no\\
T_{21}(u)\tilde{T}_{11}
+T_{22}(u)\tilde{T}_{21}(u)+(2u+1)\tilde{T}_{23}T_{31}(u)&=&
-2u\tilde{T}_{31}(u)T_{23}(u),\no\\
T_{21}(u)\tilde{T}_{12}(u)+T_{22}(u)\tilde{T}_{22}
+(2u+1)T_{23}(u)\tilde{T}_{31}(u)&=&
\tilde{T}_{31}T_{13}(u)
-(2u-1)\tilde{T}_{32}T_{23}(u)
+\tilde{T}_{33}T_{33}(u),
\eea
which come from a variant of the (graded) Yang-Baxter algebra
(\ref{rtt-ttr}) with the $R$ matrix (\ref {r1}),
\beq
\stackrel {1}{T}(u)R(2u)\stackrel {2}{\tilde{T}}(u)=
\stackrel {2}{\tilde{T}}(u)R(2u)\stackrel {1}{T}(u).
\eeq

Implementing the change $u \rightarrow u+\frac {1}{2}$ with respect to the
original problem,
one may check that these boundary K matrices satisfy the reflection equations
for the reduced problem. After some algebra, the reduced transfer matrix
$ \tau ^{(1)}(u)$ may be recognized as that for the $N$-site inhomogeneous
XXX  spin-$\frac {1}{2}$ open chain with two impurities of arbitrary spin on the boundaries,
which may be diagonalized following Ref.\cite {Skl88}.
Here we merely
give the final result,
\bea
\Lambda ^{(1)}(u;\{ u_j \}) &=&
\frac {(c_b+u-s_b)}{(c_b+u+s_b)}
\frac {(c_b+u+s_b+1)}{(c_b+u-s_b-1)}
\prod _{\a =a,b} \frac {u-c_\a-s_\a}{u+c_\a+s_\a+1}
\{ \frac {2u-1}{2u} \prod _{m=1}^M \frac 
{(u-v_m+\frac {3}{2})(u+v_m+\frac {1}{2})}
{(u-v_m+\frac {1}{2})(u+v_m-\frac {1}{2})}\no\\
& &+\frac {2u+1}{2u}
\prod _{\a =a,b} \frac {(u-c_\a+s_\a )}{(u-c_\a-s_\a)}
\frac {(u+c_\a +s_\a)}{(u+c_\a-s_\a)}
\prod _{j=1}^N \frac {(u-u_j)(u+u_j+1)}
{(u-u_j-1)(u+u_j)}
\prod ^{M}_{m=1} \frac {(u-v_m-\frac {1}{2})(u+v_m-\frac {3}{2})}
{(u-v_m+\frac {1}{2})(u+v_m-\frac {1}{2})}\},\no\\
\eea
provided the parameters $\{ v_m \}$ satisfy 
\beq
\prod _{\a=a,b}
\frac {(v_m+c_\a -s_\a-\frac{1}{2})(v_m-c_\a -s_\a-\frac{1}{2})}
{(v_m+c_\a+s_\a-\frac{1}{2})(v_m-c_\a+s_\a-\frac{1}{2})}
\prod ^N_{j=1} \frac {(v_m-u_j-\frac {3}{2})(v_m+u_j-\frac {1}{2})}
{(v_m-u_j-\frac {1}{2})(v_m+u_j+\frac {1}{2})}
=\prod ^{M}_{\stackrel {k=1}{k \neq m}} \frac {(v_m-v_k-1)
(v_m+v_k-2)} {(v_m-v_k+1)
(v_m+v_k)}. \label {bethe21}
\eeq
After a shift of the parameters $u_j \rightarrow u_j-\frac {1}{2},
v_m \rightarrow v_m + \frac {1}{2}$,the Bethe ansatz equations (\ref
{bethe11}) and 
(\ref {bethe21}) may be rewritten as follows
\bea
(\frac {u_j+\frac {1}{2}}{u_j-\frac {1}{2}})^{2L}
\prod_{\a =a,b}
\frac{u_j+c_\a +s_\a+\frac{1}{2}}{u_j-c_\a -s_\a-\frac{1}{2}}
& = & \prod_{m=1}^M \frac{u_j-v_m+\frac {1}{2}}{u_j-v_m-\frac {1}{2}}
\frac{u_j+v_m+\frac {1}{2}}{u_j+v_m-\frac {1}{2}},\no\\
\prod_{\a =a,b}\frac{v_m -c_\a-s_\a} {v_m -c_\a+s_\a}
\frac{v_m +c_\a-s_\a} {v_m +c_\a+s_\a}
\prod_{j=1}^N \frac{(v_m - u_j -\frac {1}{2})}{(v_m-u_j +\frac {1}{2})}
\frac {(v_m + u_j -\frac {1}{2})}{(v_m+u_j +\frac {1}{2})}
   &=&\prod _{\stackrel {k=1}{k \neq m}}
   \frac {(v_m-v_k-1)}{(v_m-v_k+1)}
   \frac {(v_m+v_k-1)}{(v_m+v_k+1)}
  ,\nonumber    
\eea
with the corresponding energy eigenvalue $E$ of the model 
\beq
E=-\sum ^N_{j=1} \frac {1}{u_j^2-\frac {1}{4}}.
\eeq
It should be pointed out that when $s_\a = \frac {1}{2}$,the above
results reduce to those obtained in Ref.\cite {ZG},which in turn provides
an algebraic interpretation for the applicability of the coordinate
Bethe ansatz method \cite {PW97}.

We now perform the algebraic Bethe ansatz procedure for the couplings
(\ref{con2}, \ref{con4}).   
We introduce the `doubled' monodromy matrix $U(u)$
\beq
U(u)=T(u)K_-(u){\tilde T}(u) \equiv
 \left ( \begin {array}
{ccc}
{\cal A}_{11}(u)&{\cal A}_{12}(u)&{\cal B}_1(u)\\
{\cal A}_{21}(u)&{\cal A}_{22}(u)&{\cal B}_2(u)\\
{\cal C}_1(u)&{\cal C}_2 (u)& {\cal D}(u)
\end {array} \right ), \label{u2}
\eeq
where $\tilde {T}(u)=T^{-1}(-u)$.
Substituting into the reflection equation (\ref {REs}),   
we find the following commutation relations
\bea
{\check {\cal A}}_{bd}(u_1){\cal C}_c(u_2)&=&\frac {(u_1-u_2+1)(u_1+u_2+2)}
{(u_1-u_2)(u_1+u_2+1)}r(u_1+u_2+1)^{eb}_{gh}r(u_1-u_2)^{ih}_{cd}
{\cal C}_e(u_2){\check {\cal A}}_{gi}(u_1)+\no\\
& &\frac {4(u_1+1)u_2}{(u_1+u_2+1)(2u_1+1)(2u_2+1)}r(2u_1+1)^{gb}_{cd}
{\cal C}_g(u_1) {\cal D}(u_2) - \no\\
& & \frac {2(u_1+1)}{(u_1-u_2)(2u_1+1)}
r(2u_1+1)^{gb}_{id} {\cal C}_g (u_1) {\check {\cal A}}_{ic}(u_2),\label
{cr2}\\
{\cal D}(u_1){\cal C}_b(u_2) 
&=&\frac {(u_1-u_2-1)(u_1+u_2)}
{(u_1-u_2)(u_1+u_2+1)}
{\cal C}_b(u_2){\cal D}(u_1)+\frac {2u_2}{(u_1-u_2)(2u_2+1)}
{\cal C}_b(u_1){\cal D}(u_2)\no\\
& &  -\frac {1}{u_1+u_2+1}{\cal C}_d(u_1){\check {\cal A}}_{db}(u_2).
\eea
Here $ {\cal A} _{bd}(u) = {\check {\cal A}}_{bd}(u) + \frac {1}{2u+1}
\delta _{bd}{\cal D}(u)$ and the matrix $r(u)$, which in turn satisfies the
quantum Yang-Baxter
equation, takes the form,
\beq
r^{bb}_{bb}(u)=1,~~~~~r^{bd}_{bd}=\frac
{1}{u+1},~~~~~r^{bd}_{db}(u)=\frac {u}{u+1},(b \neq d,b,d = 1,2).
\eeq
 Choosing the Bethe state $|\Omega \rangle $ as
\beq
|\Omega \rangle = {\cal C}_{i_1}(u_1) \cdots {\cal
C}_{i_N}(u_N)|\Psi\rangle F^{i_1\cdots i_N},
\eeq
with $|\Psi\rangle $ being the pseudovacuum, and applying the transfer
matrix $\tau (u)$
to the state $|\Omega\rangle$, we have
$\tau (u) |\Omega \rangle =\Lambda(u) |\Omega \rangle$, with the
eigenvalue
\bea
\Lambda (u)&=& \frac {2u+3}{2u+1}\frac {(c_b-u+s_b)}
{(c_b-u+s_b-1)}
\frac {(c_b-u-s_b-1)}
{(c_b-u-s_b-2)}
(-\frac {u+1}{u-1})^L
\prod ^N_{j=1} \frac {(u+u_j)(u-u_j-1)}{(u-u_j)(u+u_j+1)}\no\\
& &-\frac {2u}{2u+1} (-\frac {u^2}{u^2-1})^L 
\prod ^N_{j=1} \frac {(u+u_j+2)(u-u_j+1)}{(u-u_j)(u+u_j+1)}
\Lambda ^{(1)}(u;\{u_i\}),
\eea
provided the parameters $\{ u_j\}$
satisfy
\beq
\frac {2u_j+3}{2u_j+2}
\frac {(c_b-u_j+s_b)}
{(c_b-u_j+s_b-1)}
\frac {(c_b-u_j-s_b-1)}
{(c_b-u_j-s_b-2)}
(\frac {u_j+1}{u_j})^{2L}
\prod ^{M}_{\stackrel {i=1}{i \neq j}}
\frac {(u_j+u_i) (u_j-u_i-1)}
{(u_j+u_i+2) (u_j-u_i+1)}=-\Lambda
^{(1)}(u_j;\{u_i\}). \label {bethe12}
\eeq
Here $\Lambda ^{(1)}(u;\{u_i\})$ is the eigenvalue of the transfer
matrix $\tau ^{(1)}(u)$ for the reduced problem, which arises out of the
$r(u)$ matrices from the first term in the right hand side of (\ref {cr2}),
with the reduced boundary K matrices $K_{\pm}^{(1)}(u)$ as
\beq
K^{(1)}_-(u)=
  \left ( \begin {array}
{cc}
A^{(1)}_-(u)&B^{(1)}_-(u)\\
C^{(1)}_-(u)&D^{(1)}_-(u)
\end {array} \right ),\nonumber  
\eeq
where
\bea
A^{(1)}_-(u)&=&\frac {c_a^{2}-u^2-2u-s_a(s_a+1)-1-(2u+1) {\bf S}^z_a}
{(c_a-u+s_a)(c_a-u-s_a-1)},\no\\
B^{(1)}_-(u)&=&-\frac {(2u+1) {\bf S}^-_a}
{(c_a-u+s_a)(c_a-u-s_a-1)},\no\\
C^{(1)}_-(u)&=&-\frac {(2u+1) {\bf S}^+_a}
{(c_a-u+s_a)(c_a-u-s_a-1)},\no\\
D^{(1)}_-(u)&=&\frac {c_a^{2}-u^2-2u-s_a(s_a+1)-1+(2u+1) {\bf S}^z_a}
{(c_a-u+s_a)(c_a-u-s_a-1)},
\eea
and
\beq
K^{(1)}_+(u)=
  \left ( \begin {array}
{cc}
A^{(1)}_+(u)&B^{(1)}_+(u)\\
C^{(1)}_+(u)&D^{(1)}_+(u)
\end {array} \right ),  
\eeq
where
\bea
  A^{(1)}_+(u)&=&-\frac {c_b^2-u^2-2u-s_b(s_b+1)-1 -(2u+3) {\bf S}^z_b}
  {(c_b-u+s_b-1)(c_b-u-s_b-2)},\no\\
  B^{(1)}_+(u)&=&\frac {(2u+3){\bf S}^-_b}
  {(c_b-u+s_b-1)(c_b-u-s_b-2)},\no\\
  C^{(1)}_+(u)&=&\frac {(2u+3){\bf S}^+_b}
  {(c_b-u+s_b-1)(c_b-u-s_b-2)},\no\\
  D^{(1)}_+(u)&=&-\frac {c_b^2-u^2-2u-s_b(s_b+1)-1 +(2u+3) {\bf S}^z_b}
  {(c_b-u+s_b-1)(c_b-u-s_b-2)},
\eea
Here $K^{(1)}_- (u)$ ,the boundary K matrices after  the first
nesting, follows from the relations,
\bea
{\check {\cal A}}_{dd}(u)|\Psi\rangle&\equiv&
\frac {2u}{2u+1}K^{(1)}_{dd}(u)|\Psi \rangle  =
(K_-(u)_{dd}-\frac{1}{2u+1})
(-\frac{u^2}{u^2-1})^L|\Psi\rangle,\no\\
{\check {\cal A}}_{db}(u)|\Psi\rangle&\equiv&
\frac {2u}{2u+1}K^{(1)}_{db}(u)|\Psi \rangle =
K_-(u)_{db} (-\frac{u^2}{u^2-1})^L|\Psi\rangle
\eea
Indeed,applying the monodromy matrix
$T(u)$ and its ``adjoint'' ${\tilde T}(u)$ to the pseudovacuum,we have
\bea
T_{dd}(u)|\Psi\rangle&=&u^L|\Psi\rangle,~~~
T_{33}(u)|\Psi\rangle=(u+1)^L|\Psi\rangle,\no\\
T_{3d}(u)|\Psi\rangle&\neq& 0,~~~
T_{db}(u)|\Psi\rangle=0,~~~
T_{d3}(u)|\Psi\rangle= 0,\no\\
\tilde{T}_{dd}(u)|\Psi\rangle&=&(-\frac{u}{u^2-1})^L|\Psi\rangle,~~~
\tilde{T}_{33}(u)|\Psi\rangle=(-\frac{u+1}{u^2-1})^L|\Psi\rangle,\no\\
\tilde{T}_{3d}(u)|\Psi\rangle&\neq& 0,~~~
\tilde{T}_{db}(u)|\Psi\rangle=0,~~~
\tilde{T}_{d3}(u)|\Psi\rangle=0,
\eea
Then we have 
\bea
{\cal D}(u)|\Psi\rangle&=&
(-\frac{u+1}{u-1})^L|\Psi\rangle,\no\\
{\cal B}_d(u)|\Psi\rangle&=&0,~~~~~~
{\cal C}_d(u)|\Psi\rangle\neq 0,\no\\
{\cal A}_{db}(u)|\Psi\rangle&=&
(-\frac{u^2}{u^2-1})^L K_-(u)_{db}|\Psi\rangle,\no\\
{\cal A}_{dd}(u)|\Psi\rangle&=&
(-\frac {u^2}{u^2-1})^L (K_-(u)_{dd}-\frac{1}{2u+1}) |\Psi\rangle
+\frac{1}{2u+1}(-\frac{u+1}{u-1})^L|\Psi\rangle
\eea
In our calculation,use have also been made of the following 
relations,
\bea
(2u+1)T_{13}(u)\tilde{T}_{31}(u)+T_{11}(u)\tilde{T}_{11}+
T_{12}(u)\tilde{T}_{21}(u)&=&(2u+1)\tilde{T}_{31}T_{13}(u)
+\tilde{T}_{32}T_{23}(u)
+\tilde{T}_{33}T_{33}(u),\no\\
(2u+1)T_{13}(u)\tilde{T}_{32}(u)+T_{11}(u)\tilde{T}_{12}
+T_{12}(u)\tilde{T}_{22}(u)&=&2u\tilde{T}_{32}T_{13}(u),\no\\
T_{21}(u)\tilde{T}_{11}
+T_{22}(u)\tilde{T}_{21}(u)+(2u+1)\tilde{T}_{23}T_{31}(u)&=&
2u\tilde{T}_{31}(u)T_{23}(u),\no\\
T_{21}(u)\tilde{T}_{12}(u)+T_{22}(u)\tilde{T}_{22}
+(2u+1)T_{23}(u)\tilde{T}_{31}(u)&=&
\tilde{T}_{31}T_{13}(u)
+(2u+1)\tilde{T}_{32}T_{23}(u)
+\tilde{T}_{33}T_{33}(u)
\eea
which come from a variant of the (graded) Yang-Baxter algebra
(\ref{rtt-ttr}) with the quantum $R$ matrix (\ref{r2}),
\beq
\stackrel {1}{T}(u)R(2u)\stackrel {2}{\tilde{T}}(u)=
\stackrel {2}{\tilde{T}}(u)R(2u)\stackrel {1}{T}(u).
\eeq

Implementing the change $u \rightarrow u+\frac {1}{2}$ with respect to the
original problem,
one may check that these boundary K matrices satisfy the reflection equations
for the reduced problem. After some algebra, the reduced transfer matrix
$ \tau ^{(1)}(u)$ may be recognized as that for the $N$-site inhomogeneous
XXX  spin-$\frac {1}{2}$ open chain with two impurities of arbitrary spin on the boundaries,
which may be diagonalized following Ref.\cite {Skl88}.
Here we merely
give the final result,
\bea
\Lambda ^{(1)}(u;\{ u_j \}) &=&-
\frac {(c_b-u+s_b)}
{(c_b-u+s_b-1)}
\frac {(c_b-u-s_b-1)}
{(c_b-u-s_b-2)}
\prod _{\a =a,b} \frac {u+c_\a+s_\a+1}{u-c_\a-s_\a}\no\\
& &\{ \frac {2u+3}{2u+2} \prod _{m=1}^M \frac 
{(u-v_m-1)(u+v_m+1)}
{(u-v_m)(u+v_m+2)}+\frac{2u+1}{2u+2}
\prod _{\a =a,b} \frac {(u-c_\a -s_\a+1)}{(u-c_\a+s_\a+1)}
 \frac {(u+c_\a -s_\a+1)}{(u+c_\a+s_\a+1)}\no\\
& &\prod _{j=1}^N \frac {(u-u_j)(u+u_j+1)}
{(u-u_j+1)(u+u_j+2)}
\prod ^{M}_{m=1} \frac {(u-v_m+1)(u+v_m+3)}
{(u-v_m)(u+v_m+2)}\},
\eea
provided the parameters $\{ v_m \}$ satisfy 
\beq
\prod _{\a=a,b} \frac {(v_m+c_\a-s_\a+1 )(v_m-c_\a -s_\a+1)}
{(v_m+c_\a+s_\a+1)(v_m-c_\a+s_\a+1)}
\prod ^N_{j=1} \frac {(v_m-u_j)(v_m+u_j+1)}
{(v_m-u_j+1)(v_m+u_j+2)}=\prod ^{M}_{\stackrel {k=1}{k \neq m}}
\frac {(v_m-v_k-1) (v_m+v_k+1)}
{(v_m-v_k+1)
(v_m+v_k+3)}. \label {bethe22}
\eeq
After a shift of the parameters $u_j \rightarrow u_j-\frac {1}{2},
v_m \rightarrow v_m -1$,the Bethe ansatz equations (\ref
{bethe12}) and 
(\ref {bethe22}) may be rewritten as follows
\bea
(\frac {u_j+\frac {1}{2}}{u_j-\frac {1}{2}})^{2L}
\prod_{\a =a,b} 
\frac{u_j-c_\a -s_\a-\frac{1}{2}}{u_j+c_\a +s_\a+\frac{1}{2}}
\prod ^{N}_{\stackrel {i=1}{i \neq j}}
\frac {(u_j-u_i-1) (u_j+u_i-1)}
{(u_j-u_i+1)(u_j+u_i+1)}
& = & \prod_{m=1}^M \frac{u_j-v_m-\frac {1}{2}}{u_j-v_m+\frac {1}{2}}
\frac{u_j+v_m-\frac {1}{2}}{u_j+v_m+\frac {1}{2}},\no\\
\prod_{\a =a,b}\frac{v_m -c_\a-s_\a}
{v_m -c_\a+s_\a}
\frac{v_m +c_\a-s_\a}
{v_m +c_\a+s_\a}
\prod_{j=1}^N \frac{(v_m - u_j -\frac {1}{2})}{(v_m-u_j +\frac {1}{2})}
\frac {(v_m + u_j -\frac {1}{2})}{(v_m+u_j +\frac {1}{2})}
   &=&\prod _{\stackrel {k=1}{k \neq m}}
   \frac {(v_m-v_k-1)}{(v_m-v_k+1)}
   \frac {(v_m+v_k-1)}{(v_m+v_k+1)}
  ,\nonumber   
\eea
with the corresponding energy eigenvalue $E$ of the model 
\beq
E=-\sum ^N_{j=1} \frac {1}{u_j^2-\frac {1}{4}}.
\eeq

\sect{Conclusion \label{con}}

In this paper, we have studied an integrable Kondo problem describing two
impurities coupled to the one-dimensional $t-J$ open chain for specific
couplings. 
The  quantum integrability of the system follows from the fact
that the Hamiltonian may be derived from  
a one-parameter family of commuting transfer matrices. Moreover, the Bethe
Ansatz equations are obtained by means of the algebraic Bethe ansatz
approach. It should be emphasized that the boundary $K$ matrices found here
are highly nontrivial, since they can not be factorized into the product
of a c-number $K$ matrix and the local monodromy matrices. However, it is
still possible to introduce a ``singular'' local monodromy matrix $\tilde
L(u)$
and express the boundary $K$ 
matrix $K_-(u)$ as
\beq
K_-(u)=\tilde {L}(u){\tilde {L}}^{-1}(-u),
\eeq
where
\beq
\tilde L (u) =
 \left ( \begin {array}
{ccc}
u-c_a-1 -{\bf S}^z_\a&-{\bf S}^-_\a&0\\
-{\bf S}^+_\a&u-c_a-1+{\bf S}^z_\a&0\\
0&0&\e
\end {array} \right ),\label{tl}
\eeq
which constitutes a realization of the (graded) Yang-Baxter algebra (\ref
{rtt-ttr}) when $\e$
tends to $0$.
The implication of such a singular factorization deserves further
investigation.
Indeed, this implies that integrable Kondo impurities discussed here
appear to be, in some sense, related to a singular realization of
the (graded) Yang-Baxter algebra, which in turn  
reflects a hidden six-vertex $XXX$
symmetry in the original quantum $R$-matrix. Therefore, one may expect that
the formalism presented here may
be applied to other physically interesting strongly correlated
electron systems, such as the supersymmetric extended Hubbard model \cite{eks} 
and the supersymmetric U model \cite{bglz}. However,our construction
is not applicable to the one-dimensional Hubbard model and the
one-dimensional Bariev model,although a hidden six-vertex symmetry
occurs in these two physically interesting strongly correlated electron
systems \cite {MR}. Moreover,the singularity of the local monodromy
matrix (\ref {tl}) implies that we can not apply it to construct
a closed $t-J$ chain interacting with integrable Kondo-like impurities.
This is different from the conclusion by Zvyagin and Schlottmann
\cite {SZ,Zvy}, who claimed that integrable magnetic impurities exist
in the closed $t-J$ and Hubbard chains.

As shown in \cite {WV}, one can put integrable Kondo impurities on the
boundaries of the $\delta$-function interaction electron gas. Obviously,
there should be no problem in applying our construction to this model.
Another question is to extend the present construction to the
$q$-deformed case. This will lead us to integrable anisotropic Kondo impurities
coupled to the $q$-deformed version of the $t-J$ open chain.

In concluding, we would like to point out that it will be interesting to
carry out the calculation of thermodynamic equilibrium properties of
the model under consideration, based on the Bethe ansatz equations
presented here. Especially, it is desirable to calculate the finite-size
spectrum analytically, which, together with the predictions of the
boundary conformal field thery, will allow us to draw various critical
behaviour properties. Also, our construction may shed new  
light on a long-standing
problem about the quantum integrability of the conventional Kondo
model by QISM, given it has been solved using the coordinate Bethe
ansatz for a long time \cite {Wie83,And83}. 

\vskip.3in
 
One of us (HQZ) would like to express his sincere thanks to Professor
Fu-Cho Pu for encouragement and informative correspondence.
This work is supported by OPRS and UQPRS. 
JL is supported by an Australian Postdoctoral Research Fellowship.

\appendix

\sect{Derivation of the non-c-number boundary K-matrices}

In this appendix, we sketch the procedure of solving the (${\bf Z}_2$-graded) 
RE for $K_-(u)$ . To describe integrable Kondo impurites coupled
with the one-dimensional $t-J$ open chain,it is reasonable to assume
that
\beq
K_-(u) =\left (
\begin {array} {ccc}
A(u)  &B(u)&0\\
C(u)&D(u) &0\\
0&0&1
\end {array}  \right ).
\eeq
Throughout, we have omitted all the subscrips for brevity,  
reflecting  that the fermionic degrees of freedom do not occur, as it
should be for a magnetic impurity.
For the $R$-matrix (\ref{r1}), one may get from the RE (\ref {REs})
33 functional equations, of which 11 are identities. After some
algebraic analysis, together with the $su(2)$ symmetry,
we may assume that
\bea
A(u)&=&\a(u)+\b(u){\bf S}^z, ~~~B(u)=\b(u){\bf
S}^-,\no\\
C(u)&=&\b(u){\bf S}^+,~~~
D(u)=\a(u)-\b(u){\bf S}^z. \label {AK}
\eea
There are two equations automatically satisfied, leaving only 20 equations
left to be solved
\bea
& &A(u_1)B(u_2)+B(u_1)D(u_2)=
A(u_2)B(u_1)+B(u_2)D(u_1),\no\\
& &C(u_1)A(u_2)+D(u_1)C(u_2)=
C(u_2)A(u_1)+D(u_2)C(u_1),\no\\
& &u_-(A(u_1)B(u_2)+B(u_1)D(u_2))=
u_+(B(u_1)-B(u_2)),\no\\
& &u_-(A(u_2)B(u_1)+B(u_2)D(u_1))=
u_+(B(u_1)-B(u_2)),\no\\
& &u_-(C(u_1)A(u_2)+D(u_1)C(u_2))=
u_+(C(u_1)-C(u_2)),\no\\
& &u_-(C(u_2)A(u_1)+D(u_2)C(u_1))=
u_+(C(u_1)-C(u_2)),\no\\
& &u_-(A(u_1)A(u_2)+B(u_1)C(u_2)-1)=
u_+(A(u_1)-A(u_2)),\no\\
& &u_-(A(u_2)A(u_1)+B(u_2)C(u_1)-1)=
u_+(A(u_1)-A(u_2)),\no\\
& &u_-(C(u_1)B(u_2)+D(u_1)D(u_2)-1)=
u_+(D(u_1)-D(u_2)),\no\\
& &u_-(C(u_2)B(u_1)+D(u_2)D(u_1)-1)=
u_+(D(u_1)-D(u_2)),\no\\
& &u_-((u_+-1)B(u_1)D(u_2)-A(u_1)B(u_2))=
u_+((u_--1)D(u_2)B(u_1)+D(u_1)B(u_2)),\no\\
& &u_-((u_+-1)C(u_1)A(u_2)-D(u_1)C(u_2))=
u_+((u_--1)A(u_2)C(u_1)+A(u_1)C(u_2)),\no\\
& &u_-((u_+-1)A(u_2)B(u_1)-B(u_2)D(u_1))=
u_+((u_--1)B(u_1)A(u_2)+B(u_2)A(u_1)),\no\\
& &u_-((u_+-1)D(u_2)C(u_1)-C(u_2)A(u_1))=
u_+((u_--1)C(u_1)D(u_2)+C(u_2)D(u_1)),\no\\
& &u_-(A(u_1)A(u_2)-(u_+-1)B(u_1)C(u_2)
+(u_+-1)C(u_2)B(u_1)-D(u_2)D(u_1))=
u_+(D(u_2)A(u_1)-D(u_1)A(u_2))\no\\
& &u_-(A(u_2)A(u_1)-(u_+-1)B(u_2)C(u_1)
+(u_+-1)C(u_1)B(u_2)-D(u_1)D(u_2))=
u_+(A(u_1)D(u_2)-A(u_2)D(u_1)).\no\\
& &A(u_1)B(u_2)+u_+u_-D(u_1)B(u_2)+(u_--1)A(u_2)B(u_1)=
(u_+-1)(B(u_1)D(u_2)+(u_--1)B(u_2)D(u_1)),\no\\
& &D(u_1)C(u_2)+u_+u_-A(u_1)C(u_2)+(u_--1)D(u_2)C(u_1)=
(u_+-1)(C(u_1)A(u_2)+(u_--1)C(u_2)A(u_1)),\no\\
& &B(u_2)D(u_1)+u_+u_-B(u_2)A(u_1)+(u_--1)B(u_1)D(u_2)=
(u_+-1)(A(u_2)B(u_1)+(u_--1)A(u_1)B(u_2)),\no\\
& &C(u_2)A(u_1)+u_+u_-C(u_2)D(u_1)+(u_--1)C(u_1)A(u_2)=
(u_+-1)(D(u_2)C(u_1)+(u_--1)D(u_1)C(u_2)),\nonumber
\eea
with $u_+=u_1+u_2,u_-=u_1-u_2$.
Substituting (\ref {AK}) into these equations,
we find that all these equations are
reduced to the following three equations
\bea
u_+(\a(u_1)-\a(u_2))&=&u_-(-1+\a(u_1)\a(u_2)+s(s+1)\b(u_1)\b(u_2)),\no\\
u_+(\b(u_1)-\b(u_2))&=&u_-(\a(u_1)\b(u_2)+\a(u_2)\b_(u_1)-\b(u_1)\b(u_2)),\no\\
u_+(\a(u_2)\b(u_1)-\a(u_1)\b(u_2))&=& u_-(\a(u_1)\b(u_2)+\a(u_2)\b(u_1))+
u_-(u_+-1)\b(u_1)\b(u_2)).
\eea
Taking the limit $u_1 \rightarrow u_2$,these equations become
\bea
\frac{d\a(u)}{du}&=&\frac{1}{2u}(-1+{\a(u)}^2+s(s+1){\b(u)}^2),\no\\
\frac{d\b(u)}{du}&=&\frac{1}{2u}(2\a(u)\b(u)-{\b(u)}^2),\no\\
\a(u)\frac{d\b(u)}{du}-\b(u)\frac{d\a(u)}{du}&=&
\frac{1}{2u}(2\a(u)\b(u)+(2u-1){\b(u)}^2).
\label{A1}
\eea
Solving the first two equations, we have
\beq
\a(u)=\frac{(c_1c_2-u^2)(2s+1)+(c_2-c_1)u}
{(2s+1)(c_1-u)(c_2-u)},~~~~~~~~
\b(u)=\frac{2(c_2-c_1)u}
{(2s+1)(c_1-u)(c_2-u)},\label{A3}
\eeq
where $c_1$ and $c_2$ are integration constants. Substituting these 
results into the third equation in (\ref{A1}),we may establish
a relation between $c_i$:
$c_2=c_1+2s+1$.
This is nothing but the non-c-number
boundary $K$ matrix (\ref {K}) (after a redefinition of the 
constant:  
$c_1 \rightarrow -c-s-1$).

A similar construction also works for the quantum $R$ matrix (\ref
{r2}).


\end{document}